\documentclass{aa}
\usepackage{graphicx,psfig}

 \newcommand{\diff}{{\rm d}}

 \newcommand{\lp}{ \left(}
 \newcommand{\rp}{ \right)}
 \newcommand{\lc}{ \left[}
 \newcommand{\rc}{ \right]}
 
 \def\lesssim{\mathrel{\hbox{\rlap{\hbox{\lower4pt\hbox{$\sim$}}}\hbox{$<$}}}}
 \def\gtrsim{\mathrel{\hbox{\rlap{\hbox{\lower4pt\hbox{$\sim$}}}\hbox{$>$}}}}

\makeatletter
\let\@internalcite\cite
\def\cite{\@ifstar{\citeyear}{\citefull}}
\def\citefull{\def\astroncite##1##2{##1 ##2}\@internalcite}
\def\citeyear{\def\astroncite##1##2{##2}\@internalcite}
\def\citeau{\def\astroncite##1##2{##1}\@internalcite}
\def\citen{\def\astroncite##1##2{##1 (##2)}\@internalcite}
\def\possesivcite{\def\astroncite##1##2{##1's (##2)}\@internalcite}
\def\@citex[#1]#2{\if@filesw\immediate\write\@auxout{\string\citation{#2}}\fi
  \def\@citea{}\@cite{\@for\@citeb:=#2\do
    {\@citea\def\@citea{; }\@ifundefined
       {b@\@citeb}{{\bf ?}\@warning
       {Citation `\@citeb' on page \thepage \space undefined}}%
{\csname b@\@citeb\endcsname}}}{#1}}
\def\@cite#1#2{#1\if@tempswa , #2\fi}
\def\@biblabel#1{}
\makeatother
\authorrunning{Palacios et al.}
\titlerunning{Rotational mixing in Pop I low-mass stars}
\begin{document}

 \title{Rotational mixing in low-mass stars : I}
\subtitle{Effect of the $\mu$-gradients in main sequence and subgiant Pop I
  stars}

\author{Ana Palacios$^{1}$, Suzanne Talon$^{2}$, Corinne Charbonnel$^{1}$ \and Manuel Forestini$^{3}$}

\offprints{Ana Palacios ; apalacio@ast.obs-mip.fr} 

\institute{
 Laboratoire d'Astrophysique de Toulouse, CNRS UMR5572, OMP,
 14, Av. E.Belin, 31400 Toulouse, France
 \and 
 D\'epartement de Physique, Universit\'e de Montr\'eal, Montr\'eal PQ H3C 3J7,
 Canada; CERCA, 5160 Boul. D\'ecarie, Montr\'eal PQ H3X 2H9, Canada
 \and
 Laboratoire d'Astrophysique de l'Obs. de Grenoble, 414, rue de
 la Piscine, 38041 Grenoble Cedex 9, France
}

\abstract{We present a first set of results concerning stellar evolution
  of rotating low-mass stars. Our models include fully consistent transport of angular
  momentum and chemicals due to the combined action of rotation induced
  mixing (according to Maeder \& Zahn 1998) and element segregation. The
  analysis of the effects of local variations of molecular weight due to
  the meridional circulation on the transport of angular momentum and
  chemicals are under the scope of this
  study. We apply this mechanism to low mass main sequence and subgiant
  stars of population I.\\
  \indent We show that the so-called $\mu$-currents are of major importance in
  setting the shape of the rotation profile, specially near the core. 
  Furthermore, as shown by Talon \& Charbonnel (1998) and Charbonnel \& Talon
  (1999) using models without $\mu$-currents, we confirm that rotation-induced
  mixing in stars braked via magnetic torquing can explain the blue
  side of the Li dip, as well as the low Li abundances observed in
  subgiants even when $\mu$-currents are taken into account.
  We emphasize that $\mu$ variations are not to be neglected when treating
  rotation-induced mixing, and that they could be of great importance for
  latter evolutionary stages.
}

\maketitle

\section{Introduction}

Whereas standard stellar evolution models (allowing solely for
mixing in convective regions) represent a good tool for a zeroth
order description of stellar observations, they cannot account for
series of abundance patterns observed in various locations of the
Hertzsprung-Russell diagram. 

While some abundance anomalies are best explained in terms of 
surface features (as is the case of Am stars for example; see
Turcotte et al. 2000), others probably
involve deep mixing (as the He overabundances
in O stars; see Maeder \& Meynet 2000 for details).
Rotation induced mixing, in the form of meridional circulation,
baroclinic and shear instabilities, in
conjunction with microscopic diffusion has
been extensively used in order to reduce the discrepancies between
standard models and
observations, and has proved to be quite promising. 

However, following rotation induced mixing in stellar evolution codes is
challenging and, up to now, many of the theoretical results
from numerical computations have assumed some hypothesis that could 
be too restrictive. 
Here, we wish to further examine the role of horizontal chemical
inhomogeneities in the evolution of meridional circulation. 

As first pointed out by Mestel (1953), meridional circulation is
generated by a departure from spherical symmetry, and settles to restore it,
but in turn generates itself a new dissymmetry:
the meridian currents transport material with different mean molecular weights
from one place to another of the radiative zone, so that a non-spherical
$\mu$-distribution is set up which retroacts on the circulation,
diminishing its magnitude.
In this early description, assuming
that the radiative zone does not suffer any local turbulent
mixing and that the star remains in solid body rotation, 
this retroaction was actually shown to inhibit totally the circulation in chemically
inhomogeneous regions.\\
\indent Kippenhahn (1974) pointed out the importance of taking into account the
changes in molecular weight, particularly because a
strong $\nabla_{\mu}$ may represent a barrier that hinders mixing. In the
models he proposes to explain the Am phenomenon, Vauclair (1977) emphasizes also the
important role of the $\mu$-barrier and underlines the fact that it can
considerably delay mixing below the convective envelope of slow
rotators. This point, and more precisely the retroaction effect of the
$\mu$-currents on meridional circulation still in the case of Am stars, was
rediscussed further by Moss (1979).

This idea of the chocking of meridional circulation by mean molecular
gradients was pushed further by Zahn (1992). In the formalism he develops
the vertical velocity of meridional circulation, $U$, can be written as
the sum of two major terms, one describing the effects of rotation in a homogeneous star,
and the other accounting for the effects of chemical 
inhomogeneities on the circulation (see \S~2). In this description, he emphasizes the fact
that, due to the strong vertical stratification in stars, horizontal
motions are much favoured compared to vertical ones and that strongly
anisotropic turbulence is likely to set in (such a description is
also used by Tassoul \& Tassoul 1982). The effect of that turbulence is to
mix chemicals horizontally, thus slowly destroying the horizontal inhomogeneities
that give rise to $\mu$-currents.

Vauclair (1999) uses simplified expressions for rotational
mixing and couples them to microscopic diffusion in order to
examine their combined effect in slowly solid-body rotating
stars. After the build-up of the horizontal $\mu$ gradients that
block the $\Omega$-currents as in Mestel's description,
she describes a self-regulated process which could freeze both
meridional circulation and gravitational settling.

In the present paper, we present complete treatment of rotation induced
mixing, including meridional circulation, turbulence by shear instabilities
and microscopic diffusion. We will
describe the effects of such a combination of processes on the transport of
both angular momentum and chemical species in main sequence and subgiant
Pop I stars. Concerning the later point, we refer to the papers of Talon \&
Charbonnel (1998) and Charbonnel \& Talon (1999) (hereafter
TC98 and CT99 respectively) who investigated these
kind of processes under the scope of the so-called lithium dip problem, but did not
include the effects of the variations of the mean molecular weight on the
mixing. 

We will only consider stars that, according to observations in open
clusters, lie on the hot side of the lithium dip. Indeed, as pointed out by TC98,
other transport processes for momentum are known to act in lower mass
stars.
In particular, the rotation induced mixing we implemented here 
fails to reproduce the solar rotation profile as inferred
from helioseismology (Matias \& Zahn, 1997), a conclusion also
reached using other versions of detailed rotation induced mixing 
(Chaboyer, Demarque, \& Pinsonneault, 1995). TC98 suggested that the additional
mechanism responsible for momentum transport in the Sun could well
explain the rise of Li abundances on the red side of the Li dip, and
it is the point of view we adopt here by restricting our study to the
blue side.

We intend to apply our description of rotational induced mixing to abundance
anomalies in evolved stars (red giants) in a forthcoming study and will
therefore set here the theoretical stage that we will use in this and in
papers to come. In \S~2, we give a detailed description of the physics at
work, recalling the basis of the theory presented in Zahn (1992) and
Maeder \& Zahn
(1998). We will then (\S~3) describe the numerical method as well as
the inputs used in the models presented. In \S~4 we compare the models
including mixing to standard evolutionary tracks, and analyze in more
details the effects of the $\mu$-currents on the transport in \S~5. We
will finally compare our results to 
observations of main sequence and subgiant stars
in \S~6, before giving a summary and
conclusions in \S~7.  

\section{Rotation induced mixing}\label{sec:rot}

Meridional circulation takes place in the radiative interiors of rotating
stars where it is generated by the thermal imbalance induced by the
departure from spherical symmetry (Eddington 1925, Vogt 1926). 
Once established, this
large scale circulation generates in turn advection of 
momentum and thus favours the development of strong horizontal
diffusion through shear instability, which is the main source of 
turbulence\footnote{Baroclinic instabilities might set in as well,
but we adopt the view that they will be less efficient than
the shear. This point is discussed by Zahn (1992) and Talon \& Zahn (1997).}
(Zahn 1992, Talon \& Zahn 1997, Maeder \& Zahn 1998). On the other hand,
vertical turbulence is partly inhibited due to the high level of
stratification in this direction. 

Under these conditions of strong anisotropic turbulence, one can use the
hypothesis of the so-called ``shellular rotation'' (Zahn 1992). In this
configuration $\Omega \simeq \Omega(P)$ for differential rotation is
weak on isobars. All quantities depend solely on pressure and can be
split into a mean value and its latitudinal perturbation
\begin{equation}
f(P,\theta) = \overline{f}(P) + \tilde{f}(P)P_2(\cos \theta)
\label{eq1}
\end{equation}
where $P_2 (\cos \theta)$ is the Legendre polynomial of order two.
The combined effects of meridional circulation and horizontal turbulence
lead to the vertical transport of chemical species and angular momentum.

The transport of angular momentum obeys an advection-diffusion
equation treated in Lagrangian coordinates, namely
\begin{equation}
  \rho \frac{{\rm d} \left( r^{2}\Omega \right) }{{\rm d}
  t}=\frac{1}{5r^{2}}\frac{\partial }{\partial r}(\rho r^{4}\Omega
  Ur)+\frac{1}{r^{2}}\frac{\partial }{\partial r}\left( r^{4}\rho \nu
  _{v}\frac{\partial \Omega}{\partial r} \right), 
\label{momevol}
\end{equation}
$r$ being the radius, $\rho$ the density and $\nu_v$ the vertical
component of the turbulent viscosity. The vertical component of the meridional velocity is given by
\begin{eqnarray}
U(r) & = & \frac{P}{C_{p}\rho T g [\nabla_{\rm ad}-\nabla + \varphi/\delta
  \nabla_{\mu}]} \times  \nonumber\\
&  & \hspace{2.5cm} \left[ \frac{L}{M}(E_{\Omega }+E_{\mu}) + 
\frac{C_p T}{\delta} \frac{\partial \Theta}{\partial t} \right] 
\label{vcirc}
\end{eqnarray}
where $L$ is the luminosity, $M$ the mass, $P$ the pressure, $C_{\rm p}$ the specific
heat at constant pressure, $T$ the temperature, $\nabla _{\rm ad}$, $\nabla$ and
$\nabla_{\mu}$ the adiabatic, radiative and mean molecular weight gradients
respectively (Maeder \& Zahn 1998). $g$ is the modulus of the effective
gravity, defined by the hydrostatic equilibrium equation 
\begin{equation}
\vec{g} = \vec{\nabla}\Phi + \frac{1}{2}{\Omega}^2\vec{\nabla}s^2.
\end{equation}
$\varphi$ = $- \xi_{\mu} / \xi_{\rho}$ and $\delta$ = $ \xi_T /
\xi_{\rho}$, where $\xi_{\mu}$, $\xi_{\rho}$ and $\xi_T$ are the
coefficients in the general expression of the equation of state

\begin{eqnarray}
{\rm d} \ln P &= & \underbrace{ \left. { \frac{\partial \ln P}{\partial \ln \rho}}
\right| _{T,\mu}}_{\xi_{\rho}}{\rm d} \ln \rho \label{EOS} \\
&+ & \underbrace{ \left. { \frac{\partial \ln P}{\partial \ln T}} \right|
  _{\rho,\mu}}_{\xi_T}{\rm d} \ln T +
\underbrace{ \left. { \frac{\partial \ln P}{\partial \ln \mu}}\right|
  _{T,\rho}}_{\xi_{\mu}}{\rm d} \ln \mu . \nonumber
\end{eqnarray}

$E_{\Omega}$ and $E_{\mu}$ depend respectively on the rotation profile and
on the mean molecular weight as
{\bf \begin{eqnarray}
E_{\Omega} & = & 2
\left[1-\frac{\overline{\Omega^{2}}}{2 \pi G
    \overline{\rho}}-\frac{(\overline{\varepsilon}+\overline{\varepsilon}^{grav})}{\varepsilon_m}\right]\frac{\tilde{g}}{\overline{g}}\\
 & - & \frac{\rho_m}{\overline{\rho}} \left\{ \frac{r}{3}\frac{d}{dr}\left[H_T
       \frac{d}{dr} \lp \frac{\Theta}{\delta} \rp -\chi_{T} \lp
       \frac{\Theta}{\delta} \rp +
       \lp 1-\frac{1}{\delta} \rp \Theta
     \right]\right. \nonumber\\
 & - & \left. \frac{2H_T}{r} \left( 1+\frac{D_h}{K} \right) \lp
   \frac{\Theta}{\delta} \rp +\frac{2}{3}\Theta \right\} \nonumber\\
 & - &
 \frac{(\overline{\varepsilon}+\overline{\varepsilon}^{grav})}{\varepsilon_m} \left[H_T \frac{d}{dr} \lp \frac{\Theta}{\delta}\rp +(f_{\varepsilon}\varepsilon_T-\chi_T) \lp \frac{\Theta}{\delta} \rp \right. \nonumber\\
 & + & \left. \lp 2-f_{\varepsilon}-\frac{1}{\delta}\rp \Theta \right]
 \nonumber \label{eom}
\end{eqnarray}}
and
\begin{eqnarray}
E_{\mu} & = & \frac{\rho_m}{\overline{\rho}} \left\{ \frac{r}{3}\frac{d}{dr}\left[H_T
       \frac{d}{dr} \lp \frac{\varphi}{\delta} \Lambda  \rp -(\chi_{\mu} +
       \frac{\varphi}{\delta} \chi_{T}+\frac{\varphi}{\delta}) \Lambda
     \right] \right. \nonumber\\
 & - & \left. \frac{2H_T}{r} \lp \frac{\varphi}{\delta} \Lambda \rp \right\} \nonumber\\
 & + & \frac{(\overline{\varepsilon}+\overline{\varepsilon}^{grav})}{\varepsilon_m} \left[H_T
       \frac{d}{dr} \lp \frac{\varphi}{\delta} \Lambda \rp
       +f_{\varepsilon}(\varepsilon_T+ \frac{\varphi}{\delta} \varepsilon_{\mu})-\chi_{\mu} \right. \nonumber\\
 & - &  \left. \frac{\varphi}{\delta} (\chi_T+1)\Lambda \right],
\label{emu}
\end{eqnarray}
where $H_T$ is the temperature scale height, $K$ the thermal diffusivity
and $f_{\varepsilon} \equiv \overline{\varepsilon}/(\overline{\varepsilon} +
\overline{\varepsilon}^{\rm grav})$, with $\overline{\varepsilon}$ and
$\overline{\varepsilon}^{\rm grav}$ being respectively the mean nuclear and 
gravitational energy.

$\chi_{\mu}$ and $\varepsilon_{\mu}$ are logarithmic derivatives of the
radiative conductivity $\chi$ and the total energy $\varepsilon$ with
respect to $\mu$, while derivatives with respect to $T$ are
noted as $\chi_T$ and $\varepsilon_T$.

$\varepsilon_m(r) \equiv L(r)/M(r)$ and $\rho_m$ is the mean density inside the
considered level surface.\\ 
$D_h$ is the horizontal diffusion coefficient (see Eq. \ref{Dh}). 
$\Theta$ and $\Lambda$ are the relative variations over an isobar of
density and mean molecular weight respectively, with
\begin{equation}
\Theta = \frac{\tilde{\rho}}{\overline{\rho}} =
\frac{r^2}{3g}\frac{d{\Omega}^2}{d r} \hspace{0.7cm} {\rm and}
\hspace{0.7cm} \Lambda = \frac{\tilde{\mu}}{\mu}.
\end{equation}

We have considered non-stationarity, foreseeing papers to come where the
mixing processes described here will be further investigated inside RGB
stars. The effects of non-stationarity on the circulation, which
need to be included as soon as nuclear burning takes place in shells, appear
through the terms $\partial \Theta / \partial t$ and
$\overline{\varepsilon}^{\rm grav}$ (Maeder \& Zahn 1998).

As shown by Chaboyer \& Zahn (1992), the vertical transport of chemicals
through the combined action of vertical advection and strong
horizontal diffusion 
can be described as a pure diffusive process.
Indeed, splitting the concentration $c_i$  of the $i^{th}$
element according to Eq. (\ref{eq1}), the advective transport 
in the rigorous evolution equation
for the mean concentration
\begin{equation}
\frac{\partial \overline{c_i}}{\partial t} + \frac{1}{r^2}\frac{\partial
  r^2 \rho \left< \tilde{c_i}U \right> }{\partial r} = \frac{1}{r^2} 
\frac{\partial}{\partial r}  \left[
r^2 \rho D_v \frac{\partial \overline{c_i} }{\partial r} \right]
\label{diffelts}
\end{equation}
can be replaced by a diffusive term, with an effective
diffusion coefficient given by
\begin{equation}
D_{\rm eff} = \frac{|rU(r)|^2}{30D_h}.
\label{Deff}
\end{equation}
The vertical transport of chemicals then
obeys a diffusion equation which, in addition to this macroscopic transport,
also accounts for (vertical) turbulent transport, microscopic diffusion (see \S~3)
and nuclear reactions:  
\begin{eqnarray}
\rho  \frac{{\rm d} \overline{c_i}}{{\rm d} t} & = & \dot{c_i} +
\frac{1}{r^2}\frac{\partial}{\partial r}\left[r^2\rho U_{\rm diff}\overline{c_i}\right] \nonumber\\
 & + & \frac{1}{r^2}\frac{\partial}{\partial r}\left[r^2\rho
   \left(D_{\rm eff}+D_{v}\right)\frac{\partial \overline{c_i}}{\partial r}\right].
\label{diffelts2}
\end{eqnarray}
Microscopic diffusion appears through the velocity term ($\propto
U_{\rm diff}$), and $\dot{c_i}$ represents the variations of chemical
composition due to nuclear reactions. $D_v$ is the vertical turbulent
diffusion coefficient, taken equal to the turbulent viscosity (Zahn 1992). 
The total diffusion coefficient $D_{\rm tot}$ for chemicals 
can be written as the sum of
three coefficients: 
\begin{equation}
D_{\rm tot} = D_{\rm mic}+D_{\rm eff}+D_v.
\end{equation}

$\Lambda$ is tightly related to the horizontal inhomogeneities of chemicals, and obeys
a similar law as the concentration of the chemicals under the
assumption of shellular rotation,
\begin{equation}
\frac{\partial \Lambda}{\partial t} + U \frac{\partial
  \ln\overline{\mu}}{\partial r} = - \frac{6}{r^2}D_h \Lambda 
\label{lambda}
\end{equation}
(see Chaboyer \& Zahn 1992 and Maeder \& Zahn 1998 for details).
Even though this equation admits a stationary solution for
$t \gg r^2/6D_h$, we will follow its time evolution since it allows
for greater stability in the numerical algorithms.

Because of the low viscosity in stellar interiors, vertical shear due to
differential rotation eventually becomes turbulent, and we will assume
that this instability dominates, noting that baroclinic
instabilities may also grow.  
However,
development of the shear instability is possible only under certain conditions
which are fulfilled
when the flow satisfies to both the Reynolds and
the Richardson instability criteria. 

The Richardson criterion compares the stabilizing effect of the entropy
gradient to the amount of energy that can be extracted from
differential rotation. The instability condition is in general
written as
\begin{equation}
Ri = \frac{N^2}{\lp {\rm d}u/{\rm d}z \rp ^2} \leq Ri_{\rm crit} \simeq 
\frac{1}{4}.
\end{equation}
This criterion is extremely severe and is seldom satisfied in stellar interiors.

However, as first pointed out by Townsend (1958), thermal diffusivity 
can decrease the stabilizing effect of entropy stratification.
Maeder (1995) used this characteristic to show that in actual conditions,
there will always exist a small enough length scale that will make
the fluid unstable. The modified Richardson criterion becomes
\begin{equation}
Ri = \frac{ \lp \frac{\Gamma}{\Gamma+1} \rp N_T^2 + N_{\mu}^2 }
{\lp {\rm d}u/{\rm d}z \rp ^2} \leq Ri_{\rm crit} \label{rimod}
\end{equation}
where $\Gamma= v \ell / 6 K$, $v$ is the turbulent velocity,
$\ell$ the turbulent length scale, 
$K$ is the thermal diffusivity, $N_T^2$ is the thermal
part of the Br\"unt-V\"ais\"al\"a frequency and $N_{\mu}^2$ is related
to mean molecular weight gradients. $\Gamma$ can be viewed as
a turbulent Peclet number, the turbulent viscosity $\nu_v=v \ell$ replacing
the normal microscopic viscosity. 
The turbulent viscosity is then obtained by setting the equality
in Eq. (\ref{rimod}). Further discussion on these topics may be found
in Canuto (1998), who suggests that, in such a context, the critical
Richardson number might actually be closer to unity.

In order for the instability to grow, the associated turbulent
viscosity must also be larger than the microscopic viscosity $\nu$, as
expressed by the Reynolds criterion
\begin{equation}
\nu_v \leq \nu Re_c
\end{equation}
where $Re_c$ is the critical Reynolds number ($\simeq 10$).

The estimation of the appropriate formulation for the Richardson criterion
is critical to give a good estimation of the turbulent viscosity, and in
turn of the turbulent diffusion coefficient. Meynet \& Maeder (2000)
present a critical review of the different expressions proposed so far
concerning this coefficient. In the present work, we shall take the
expression derived by Talon \& Zahn (1997) for the vertical component of
the turbulent viscosity  
\begin{equation}
 D_v = \nu_v = \frac{8}{5} \frac {Ri_{\rm crit}  (r d
 \Omega/dr)^2}{N^{2}_{T}/(K+D_h)+N^{2}_{\mu}/D_h},
\label{Dv}
\end{equation}
which adds to Maeder's prescription by also taking into account the 
weakening effect of the strong horizontal
diffusion on the entropy stratification associated with the (vertical)
$\mu$ gradients. 
The factor $\frac{8}{5}$ includes, in addition to a geometrical factor, the
coefficient $\frac{2}{5}$ found by Maeder (1995) when deriving the
criterion for shear instabilities assuming spherical geometry for the
turbulent eddies.

$D_h$ is the horizontal diffusion coefficient and is such as $D_h \gg D_v$,
thus leading to strong anisotropic turbulence, which is one of the basic
hypothesis of the derivation presented. 
This coefficient appears directly in Eq. (\ref{lambda}) describing the
evolution of the variations of $\Lambda$, as well as in the vertical and
effective diffusivity coefficients. 
Its evaluation is the weakest point of the theory first developed by
Zahn (1992). In the present paper, we use an expression linking this
coefficient to the advection of angular momentum
\begin{equation}
D_h = \frac{rU}{C_h}\left[\frac{1}{3}\frac{{\rm d} \ln \rho r^2 U}{{\rm d}
    \ln r}-\frac{1}{2}\frac{{\rm d} \ln r^2\Omega}{{\rm d} \ln r}\right].
\label{Dh}
\end{equation}
$C_h$ is a free parameter of order unity (see Zahn 1992) which describes
the weakening effect of horizontal turbulence on the vertical transport
of chemicals. It appears through $D_h$ in both effective ($D_{\rm
  eff}$) and turbulent ($D_v$) diffusion coefficients as can be seen from
Eq. (\ref{Deff}) and (\ref{Dv}). As $D_{\rm eff} \propto C_h$ and $D_v
\propto {C_h}^{-1}$, the net effect of varying this parameter on the global
diffusion coefficient $D_{\rm tot}$ might not be very important. 

On the other hand, the magnitude of turbulent diffusion $D_v$
depends on $Ri_{\rm crit}$ (see Eq. \ref{Dv}). Changing the Richardson
criterion as previously suggested, by  using a larger critical number
$Ri_{\rm crit}$, will result in an enhancement of $D_v$ and in turn of the
global diffusion coefficient.   

\section{Numerical simulations : inputs}

 We present and discuss models computed with the Grenoble stellar
 evolution code, STAREVOL. We included the transport of angular momentum and 
 chemical species due to the combined action of meridional circulation,
 turbulence and microscopic diffusion (in the case of chemicals), following
 the formalism of Maeder \& Zahn (1998) as described in the previous section.

\subsection{Numerical treatment of the transport of angular momentum}

 The 4$^{th}$ order differential equation for the transport of angular momentum 
 (Eq.~\ref{momevol}) is
 split into four 1$^{st}$ order equations. The system is complemented by
 Eq.~(\ref{lambda}) in order to follow the feedback of variations
 of the mean molecular weight on transport. 

 We use the Newton-Raphson method according to Henyey (1964) to solve for
 angular momentum transport within the evolutionary code. 
 The first boundary conditions impose momentum conservation at
 convective boundaries
\begin{displaymath}
\begin{array}{ll}
{\displaystyle \frac{\partial}{\partial t}
\lc \Omega \int _{r_t}^R  r^4 \rho \, \diff r \rc =
-\frac{1}{5} r^4 \rho \Omega U + {\cal F}_\Omega } & ~~~{\rm for} ~ r=r_t \vspace{0.2cm}\\
{\displaystyle \frac{\partial}{\partial t}
\lc \Omega \int _0^{r_b}  r^4 \rho \, \diff r \rc = \frac{1}{5} r^4 \rho \Omega
U } & ~~~{\rm for} ~~r=r_b.
\end{array}
\end{displaymath}
We complement them by requiring the absence of differential rotation at
convective boundaries
\begin{displaymath}
\begin{array}{ll}
{\displaystyle \frac{\partial \Omega}{\partial r} }= 0 & \hspace{4.88cm}{\rm for}~ r = r_t,~r_b \\
\end{array}
\end{displaymath}
We also impose $\Lambda=0$ there, a natural consequence of the fact that convective
regions are assumed to be chemically homogeneous.
$r_t$ and $r_b$ correspond to the top (surface) and bottom (center) of the
radiative zone respectively. ${\cal F}_{\Omega}$ represents the torque
applied at the surface of the star. Note that momentum conservation is
equivalent to $U= 0$ when there is no convective core.

In the models presented here, we considered a classical perfect gas so that $\delta =
\varphi = 1$ in Eqs.~(\ref{vcirc}),~(5) and (\ref{emu}).
We also used $C_h$ = 1 and $Ri_c$ = 0.25, thus not allowing these two
parameters to vary.  

The transport of chemicals according to Eq.~(\ref{diffelts2}) is treated 
via a finite element method. Boundary conditions impose conservation of the
integrated mass of each element and can be written as follows
\begin{displaymath}
\begin{array}{lll}
{\displaystyle \frac{\partial}{\partial t}
\lc c_i \int _{r_t}^R  r^2 \rho \, \diff r \rc} & = &
{\displaystyle -r^2 \rho \left( U_{\rm diff}\overline{c_i}\right) -r^2 \rho \left( D_v +
  D_{\rm eff} \right) \frac{\partial c_i}{\partial r}}\\
& &{\displaystyle  - \dot{M}c_i } ~~~~~~~~~~~~~~~~~~~~~~~~~{\rm for} ~ r=r_t \vspace{0.2cm}\\
\end{array}
\end{displaymath}
\begin{displaymath}
\begin{array}{lll}
{\displaystyle \frac{\partial}{\partial t}
\lc c_i \int _0^{r_b}  r^2 \rho \, \diff r \rc} & = &
{\displaystyle r^2 \rho \left( U_{\rm diff}\overline{c_i}\right) + r^2 \rho \lp D_v +
D_{\rm eff} \rp \frac{\partial c_i}{\partial r} } \\
 & & ~~~~~~~~~~~~~~~~~~~~~~~~~~~~~~~~{\rm for} ~~r=r_b,\\
\end{array}
\end{displaymath}
where $\dot{M}$ is the mass loss rate.

We emphasize that rotational induced mixing is treated in a
self-consistent way and that coupling to the evolutionary code ensures 
feedback of transport mechanisms on structure at all
evolutionary phases. 

\subsection{Structure equations}
One of the basic effects of rotation on internal structure is 
to break the spherical symmetry of the star. Rigorously, corrections due to 
this departure from spherical symmetry should thus be applied when 
writing the internal structure equations.
Meynet \& Maeder (1997) developed a formalism suitable in the 
``shellular rotation'' context. 
They showed that rotation has little influence on the 
hydrostatic structure of massive stars, while the effects at the stellar surface 
(in particular on the effective temperature) may be important.
In the HR diagram, these effects may be of the same order of magnitude 
as those due to mixing. 
In the present work, we made the assumption of negligible departures from
spherical symmetry, and we did not include these corrections (i.e., stellar
structure equations are solved in a standard way). The effects on surface
parameters such as $L$ and $T_{\rm eff}$ discussed here are
thus only due to rotation-induced mixing. 
 
\subsection{Evolution of the rotational surface velocity}

As already mentioned, we only consider stars hotter than 6500 K on the ZAMS,
associated to the hot side of the lithium dip in open clusters. We refer to the
work of Gaig\'e (1993) on the Hyades F-stars to choose the initial velocity
of our models and further determine the braking to be applied according to
the temperature. For further discussion on rotation in F-stars, the
reader is referred to TC98 and CT99 where these aspects are more
extensively discussed.

In the present paper, stars have an initial velocity of 110 ${\rm
  km}.{\rm s}^{-1}$ and are assumed to undergo magnetic braking while
  arriving on main sequence (see Tab.~\ref{table1}) as
suggested by the early work of Schatzman (1962). This braking was proved to
  be more efficient for thicker convective envelopes, that is to say for
  lower masses at given metallicity, a trend which is confirmed by the
  observations. Their rotational velocity is further reduced to only a few
  ${\rm km}.{\rm s}^{-1}$ when they are on the subgiant branch due to the
  star's expansion.

 The braking law adopted follows the description of Kawaler (1988)
\begin{equation}
 \frac{{\rm d} J}{{\rm d}t} = \left\{
\begin{array}{l l }
-K \Omega^3 \lp {\displaystyle \frac{R}{R_\odot}} \rp ^{1/2} 
\lp {\displaystyle \frac{M}{M_\odot} }\rp ^{-1/2} & 
(\Omega \leq \Omega_{\rm sat}) \\
 & \\
-K \Omega \, {\Omega^2}_{\rm sat} \lp {\displaystyle \frac{R}{R_\odot}} \rp ^{1/2} 
\lp {\displaystyle \frac{M}{M_\odot} }\rp ^{-1/2}
& (\Omega > \Omega_{\rm sat}).  
 \end{array}   \right.
\end{equation}
This formulation corresponds to a field geometry intermediate between
a dipolar and a radial field (Kawaler 1988). Neglecting the evolution
of stellar structure and assuming solid body rotation, this leads to
a Skumanich law in $t^{-1/2}$ for the surface velocity.
It is widely used in the literature
(Chaboyer et al. 1995, Krishnamurti et al. 1997, Bouvier et al. 1997, Sills
\& Pinsonneault 2000). The parameter $\Omega_{\rm sat}$ expresses
the fact that magnetic field generation saturates at some critical
value, as shown by various diagnostics of magnetic activity (see
Saar 1996 and references therein). This saturation is
actually required in order to retain a sufficient amount of
fast rotators in young clusters, as originally suggested by
Stauffer \& Hartmann (1987).
The constant $K$ is related to the magnitude of the magnetic field
strength. In particular, in the mass range we are studying here,
the convection zone shrinks significantly from the $1.35 {\rm M}_{\odot}$ 
to the $1.8 {\rm M}_{\odot}$ model, and this parameter, which is
normally calibrated on the Sun and taken to be a constant in all
stars (see Bouvier et al. 1997) could very well vary here.

In studies of stars less massive than $\sim \, 1.2 {\rm M}_{\odot}$,
there is a suggestion that $\Omega_{\rm sat}$ varies with mass,
and a scaling in $\tau_{\rm conv}^{-1}$ is invoked
(Barnes \& Sofia 1996, Krishnamurti et al. 1997,
Bouvier et al. 1997, Sills \& Pinsonneault 2000). This is required to explain
the fact that K stars rotate more rapidly than G stars in the Hyades.
However, any diagnosis depends on the actual model for momentum
distribution in the stellar interior.
For the masses under the scope of this study, braking remains much smaller than
it is for lower mass stars, as reflected by the fact that at the age
of the Hyades, even the less massive model is still rotating at $30 ~{\rm km.s}^{-1}$.

Table \ref{table_rot} presents the rotational characteristics of our models,
as well as the global convective time-scale as defined by Kim \& Demarque (1996)
\begin{equation}
\tau_{\rm conv} = \int_{r_t}^R {\rm d} r / v_{\rm conv}.
\end{equation}
The scaling of $K \Omega_{\rm sat}^2$ is clearly not inversely proportional 
to $\tau_{\rm conv}$. Furthermore, braking for stars more massive
than $\sim 1.35 {\rm M}_{\odot}$ is less efficient than it is in lower
mass stars, while their convective time-scales are lower.
We are thus lead to believe it is a mass range for 
which the value of $K$ itself varies, in agreement
with our knowledge that it is also a transition region for stellar activity.

Let us note that all our calculations were performed in the saturated regime,
and that our stars follow the evolution of the mean velocity for the 3 considered
clusters ($\alpha$ Per, Pleiades, Hyades). 
It is thus not possible to disentangle the effect of a variation
of $\Omega_{\rm sat}$ and a variation of $K$.

\begin{table}
\caption{Rotational parameters of the models}
\noindent
 \begin{tabular}{c c c c c c}
 \hline
\footnotesize{M} &\footnotesize{ $K \Omega_{\rm sat}^2$} & \footnotesize{ $\tau_{\rm conv}$} &\footnotesize{$T_{\rm eff}$}
& \footnotesize{$v$} & \footnotesize{$Age$}  \\ 
\footnotesize{$({\rm M}_{\odot})$} & \footnotesize{$$} & \footnotesize{days}
& \footnotesize{$(K)$} &
\footnotesize{(km.s$^{-1}$)} \\
 \hline
 \hline
\footnotesize{1.35} & \footnotesize{$2.3 \times 10^{43}$} & 6.0 & \footnotesize{6620} & \footnotesize{70.3} & \footnotesize{$\alpha$ Per} \\
                    &                                     &     & \footnotesize{6630} & \footnotesize{43.6} & \footnotesize{Pleiades} \\
                    &                                     &     & \footnotesize{6625} & \footnotesize{30.6} & \footnotesize{Hyades} \\ \hline
\footnotesize{1.4 } & \footnotesize{$4.4 \times 10^{42}$} & 4.0 & \footnotesize{6770} & \footnotesize{97.5} & \footnotesize{$\alpha$ Per} \\
                    &                                     &     & \footnotesize{6760} & \footnotesize{66.2} & \footnotesize{Pleiades} \\
                    &                                     &     & \footnotesize{6745} & \footnotesize{46.3} & \footnotesize{Hyades} \\ \hline
\footnotesize{1.5 } & \footnotesize{$5.1 \times 10^{41}$} & 3.4 &\footnotesize{7080} & \footnotesize{100} & \footnotesize{$\alpha$ Per} \\
                    &                                     &     & \footnotesize{7045} & \footnotesize{83.0} & \footnotesize{Pleiades} \\
                    &                                     &     & \footnotesize{6990} & \footnotesize{76.0} & \footnotesize{Hyades} \\ \hline
\footnotesize{1.8 } & \footnotesize{$7.1 \times 10^{40}$} & 0.44 &\footnotesize{8280} & \footnotesize{110} & \footnotesize{$\alpha$ Per} \\
                    &                                     &     & \footnotesize{8020} & \footnotesize{94.4} & \footnotesize{Pleiades} \\
                    &                                     &     & \footnotesize{7625} & \footnotesize{84.1} & \footnotesize{Hyades} \\ \hline
\label{table_rot}  
\end{tabular}
\end{table}

\subsection{Treatment of the microscopic diffusion \label{diff_micro}}

We included microscopic diffusion in the
form of gravitational settling as well as that related to thermal
gradients, using the formulation of Paquette et
al. (1986).
Atomic diffusion for trace elements Li, Be, B must be addressed with care
in the limited range of effective temperature on the left side of the Li dip. Indeed,
Richer \& Michaud (1993) showed that, for main sequence stars with an effective
temperature in the range $\sim 6900-7100$~K, radiative forces
on Li and Be could lead to increased surface abundances, at least
if large scale mixing is not efficient enough to counter balance them.
The recent discovery of a slowly rotating Li rich dwarf
(Deliyannis et al. 2002) is nicely explained in this framework.
This had lead TC98 and CT99 to remove microscopic diffusion on
those elements, in the corresponding temperature range. 
Here, we adopt a different view point.
We compute a first series of models including microscopic diffusion
(but not radiative forces) on all light elements. Then, for stars
in the range of effective temperature between $\sim 6900-7100$~K, we
also compute models without microscopic diffusion on Li, Be and B.
The differences between these two series of models give an indication of the
amount of microscopic diffusion that remains in our rotating models.
As we shall show, this amount is minute. 
One must keep in mind though
that the real effect of microscopic diffusion in these stars 
will be to slightly enhance Li and Be abundances rather than
diminish them.\\ 
Let us note that radiative forces have a magnitude which is
similar to that of settling (see Richer et al. 2000 Figs. 5 \& 6).
Thus, if mixing is efficient enough to (almost) inhibit settling, it
will also inhibit levitation by radiation.\\
For main sequence stars with ${\rm T}_{\rm eff} >
7100$~K, radiative forces do not play a major role compared to atomic
diffusion, which is thus the only process (apart from rotation-induced
mixing) which is applied.

\subsection{Input physics}

A rather detailed description of our stellar evolution code can be found in
Siess et al. (1997) (see also Siess et al. (2000) for more recent aspects). 
As far as
main sequence and subgiant models are concerned, let us briefly summarize here
the relevant physics. Our equation of state has been developed from the Pols et
al. (1995) formalism. Thermodynamical features of each plasma component
(ions, electrons and photons, as well as $\mathrm{H}^{-}$ and $\mathrm{H}_{2}$)
are obtained by minimizing the Helmholtz free energy that includes separately
non-ideal effects (Coulomb shielding and pressure ionisation). In input,
instead of the usual pressure or density variable, it uses a new independent
variable that describes the electron degeneracy. This allows in turn to treat
analytically the ionisation process and provides very smooth profiles of
thermodynamical quantities. It is particularly well suited to treat partially
ionised and partially degenerated regions, pressure ionisation and Coulomb
interactions in a stellar evolution code. More details about this equation of
state are given in Siess et al. (2000).

Radiative opacities are interpolated from Alexander \& Fergusson (1994) at
temperatures below 8000 K, and from Iglesias \& Rogers (1996) at higher
temperatures. Atmospheres are integrated in the gray, plan-parallel and
Eddington approximations. Our nuclear reaction network allows to follow the
abundance evolution of 53 species (from ${}^{1}\mathrm{H}$ to
${}^{37}\mathrm{Cl}$) through 180 reactions. All nuclear reaction rates
have been updated with the NACRE compilation (Angulo et al., 1999).
The value of $\alpha_p = l/H_p$, the ratio of the mixing length to the 
pressure scale height, is taken equal to 1.75. No overshooting is considered 
for convection.

\section{Global characteristics of the models}
\begin{table}
\caption{Characteristics of the stellar models and rotation
  velocities. Models computed with Z = 0.02. The turn-off (TO) is
  defined as the point for which the central hydrogen mass fraction is
  lower than $10^{-9}$.}
\noindent
 \begin{tabular}{c c c c c c}
 \hline
\footnotesize{M} &\footnotesize{ $v_{\rm init}$} & \footnotesize{$v_{\rm Hyades}$}
& \footnotesize{$T_{\rm eff,TO}$} & \footnotesize{$L_{\rm TO}$} &
\footnotesize{$t_{\rm TO}$} \\ 
\footnotesize{$({\rm M}_{\odot})$} & \footnotesize{$({\rm km}.{\rm
    s}^{-1})$} & \footnotesize{$({\rm km}.{\rm s}^{-1})$} &
\footnotesize{(K)} & \footnotesize{$({\rm L}_{\odot})$}  
 & \footnotesize{(Gyrs)}\\
 \hline
 \hline
\footnotesize{1.35} & \footnotesize{110} & \footnotesize{30} &
\footnotesize{6113} & \footnotesize{5.92} & \footnotesize{4.140} \\
 & \footnotesize{0} & \footnotesize{0} &
\footnotesize{6328} & \footnotesize{5.68} & \footnotesize{3.418} \\ \hline
\footnotesize{1.4 } & \footnotesize{110} & \footnotesize{51} &
\footnotesize{6218} & \footnotesize{6.92} & \footnotesize{3.555} \\
 & \footnotesize{0} & \footnotesize{0} &
\footnotesize{6451} & \footnotesize{6.42} & \footnotesize{2.885} \\\hline
\footnotesize{1.5 } & \footnotesize{110} & \footnotesize{80} &
\footnotesize{6382} & \footnotesize{8.80} & \footnotesize{2.967}\\
 & \footnotesize{0} & \footnotesize{0} &
\footnotesize{6585} & \footnotesize{8.38} & \footnotesize{2.404}\\\hline
\footnotesize{1.8 } & \footnotesize{110} & \footnotesize{82} &
\footnotesize{7001} & \footnotesize{17.81} & \footnotesize{1.470}\\
 & \footnotesize{0} & \footnotesize{0} &
\footnotesize{7255} & \footnotesize{17.17} & \footnotesize{1.317}\\\hline
\footnotesize{2.2 } & \footnotesize{110} & \footnotesize{77} &
\footnotesize{8034} & \footnotesize{39.41} & \footnotesize{0.788}\\
 & \footnotesize{0} & \footnotesize{0} &
\footnotesize{8228} & \footnotesize{39} & \footnotesize{0.746}\\
 \hline
 \label{table1}  
\end{tabular}
\end{table}
We present evolutionary results for five different masses (1.35, 1.4, 1.5, 1.8 and 2.2
${\rm M}_{\odot}$) at Z = 0.02. Models include 
complete treatment of angular momentum and chemicals transport,
taking into account the effects of changes of $\mu$
($E_{\mu} \ne 0$) and applying microscopic diffusion to all elements. 

These models sample the effective temperature domain covered by the
observations of main-sequence stars on the hot side of the lithium dip in
young open clusters with nearly solar metallicity such as the Hyades, Coma
Berenices and Praesepe. They also span the mass range of Pop I subgiants to
which our results will be compared in a further section (\S~6).

\begin{figure}[t]
\resizebox{\hsize}{!}{\includegraphics{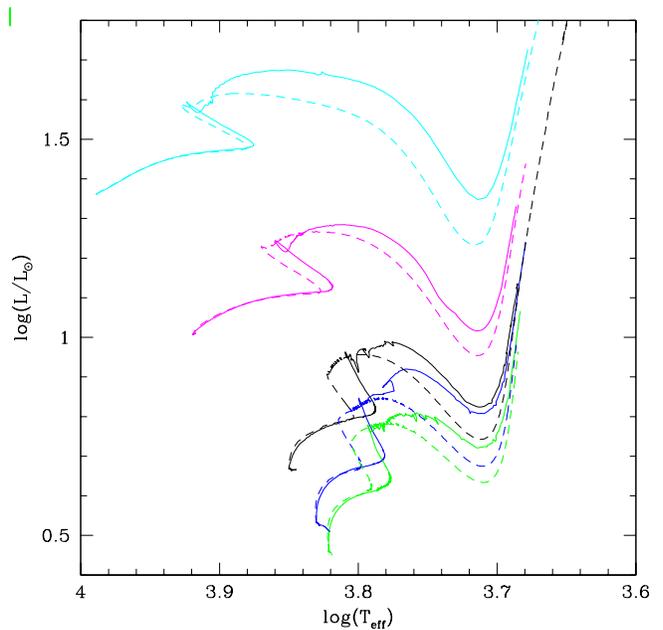}}
\caption{HR diagrams for the five masses considered in this paper. Dashed
  lines are for standard models and solid lines are for models with the
  ``complete mixing'' and $C_h$ = 1. Masses are as indicated on the
  plot. Z = 0.02 for all models presented.} 
\label{HR}
\end{figure}

In Fig.~\ref{HR}, we plot the evolutionary tracks of our rotating models in the
Hertzsprung-Russell diagram (solid lines), comparing them to their
standard analogs (dashed lines). 

Table~\ref{table1} presents some characteristics of the computed models
with and without rotation. All models were computed with the same set of
parameters for rotation induced mixing (${\rm C}_h = 1 $ and $Ri_{\rm crit} =
0.25$), but with different rotational histories (see \S~3.2).

One can see clearly how mixing affects main-sequence evolution.
In rotating models, the turn-off is moved towards cooler temperatures 
while the luminosity is slightly enhanced. 
Due to surface effects, still cooler effective temperatures 
(by up to $\sim 100~{\rm K}$ in the fastest stars considered here) would 
be obtained if the effects of rotation on structure equations 
were taken into account (Per\'ez Hern\'andez et al. 1999). 
Let us note that the actual value depends on the inclinaison angle
of the star. 

Our models are computed without convective core overshooting. 
We checked however that for the 1.8M$_{\odot}$ model, the effect of 
mixing on main sequence lifetime and on the turnoff in the HRD 
is equivalent to an overshooting distance for convection of about 0.1$H_P$
(where $H_P$ is the pressure scale height at the edge of the classical core).
This value is smaller than the one (0.2) typically used  
to fit the observed color-magnitude diagrams of young and intermediate age 
galactic open clusters
(Meynet et al. 1993, Lebreton 2000 and references therein) . Whether the rotating models would fit the cluster 
isochrones remains to be checked. This requires a more extended grid of masses 
and is out of the scope of the present paper.

On the subgiant branch, stars evolve at higher
luminosities. This is due to the effect of microscopic diffusion during
the main sequence phase. Indeed, even if it is partly balanced by
the action of large scale mixing, microscopic diffusion 
leads to the build up of a negative
helium gradient at the base of the convective envelope. 
When the first dredge-up begins, the envelope encounters regions 
with a lower opacity (due to enhancement of helium) 
leading to enhanced luminosities compared to standard models.
When reaching the RGB, differences in effective temperature and
luminosity between standard and rotating models become marginal:
both tracks are almost identical in the HR diagram (see Fig.~\ref{HR}).
   
Models with rotational mixing behave as if their metallicity was lower than
in the standard cases. This trend was already emphasized by CT99. 

On the other hand, the lifetimes of rotating models are enhanced with
respect to those of standard models. In fact, due to the positive slope of
the hydrogen profile, mixing slightly feeds the core with fresh hydrogen
fuel. The exhaustion of hydrogen in the central region is delayed and the time
spent on main sequence increases. 

 In the mass range we are studying here, the less massive the star, the
  more turn-off ages differ between rotating and non-rotating models\footnote{The
  time spent on the main sequence is increased by more than $20\%$ in the
  lower mass stars, and by $\approx 10\%$ in the higher mass models.}. This
  trend with mass has to be related to braking, which is more important for
  lower mass stars as imposed by both theory and observations. Stronger
  mixing corresponds to a more efficient extraction of angular momentum at
  the surface, thus leading to enhanced meridional circulation and mixing.
There should thus be a correlation between mixing and braking, as we 
will show from the comparison of our models with observations (see \S~6).

\begin{figure*}[t]
\resizebox{\hsize}{!}{\includegraphics[angle=-90]{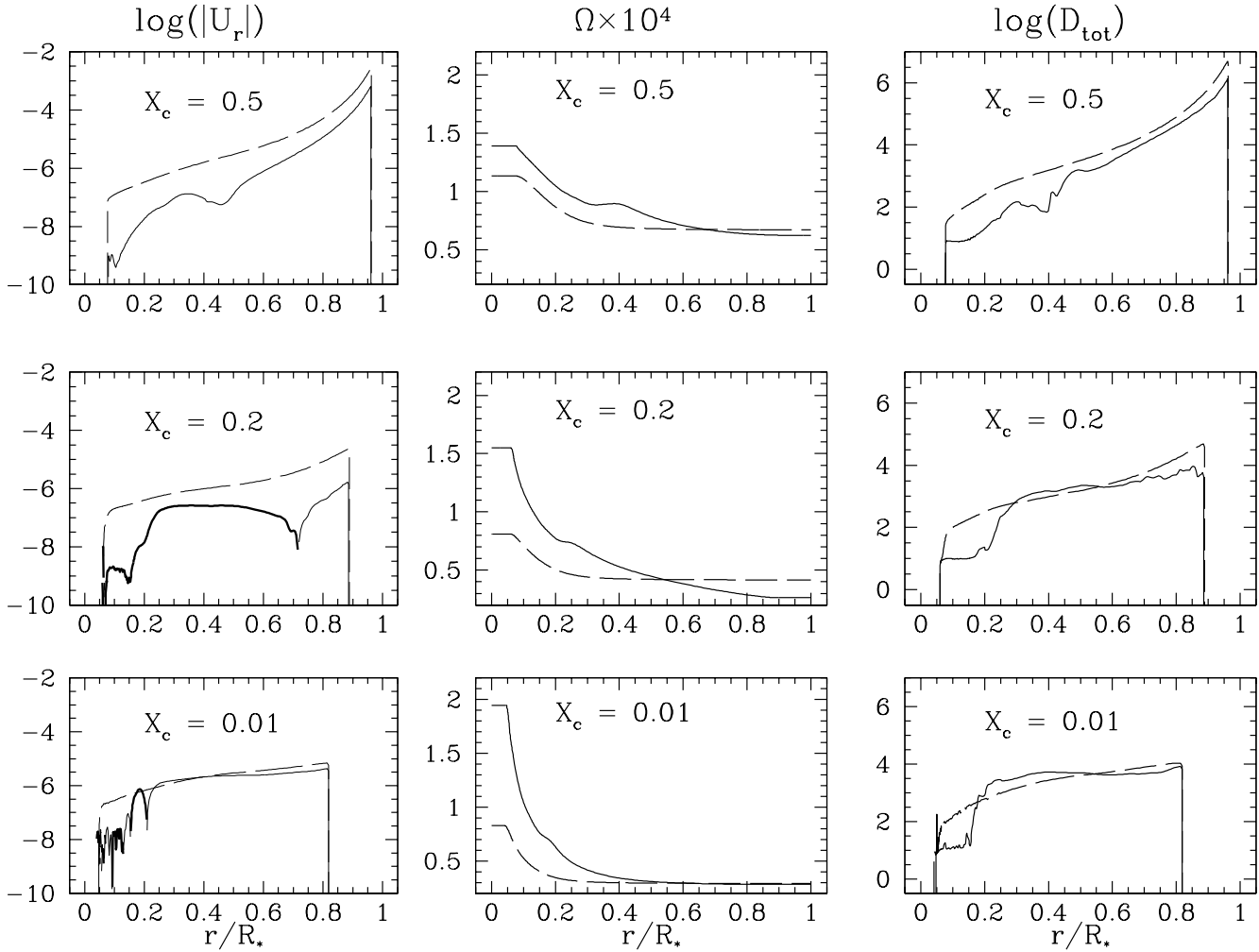}}
\caption{Profiles of the vertical component of the meridional velocity
  $U_r$ (left panels), angular velocity $\Omega$ (middle panels) and
  diffusion coefficients (right panels) inside a $1.5~{\rm M}_{\odot}$ , Z
  = 0.02 star. Solid lines are for the model with $E_{\mu}
  \neq 0$ and dashed lines for the model with $E_{\mu} = 0$. Central
  hydrogen content is as indicated. In the left panels, bold parts in the solid lines indicate
  positive values of $U_r$.} 
\label{Ur}
\end{figure*}

\section{Effects of $\nabla_{\mu}$ and ${\mu}$-currents}  

  To discuss the effects of chemical composition variations 
  on transport, we will focus on the $1.5~{\rm M}_{\odot}$
  model. We present a comparative study of the results obtained
  with and without $\mu$-currents ($E_{\mu} = 0$ or $E_{\mu} \ne 0$. In both cases all parameters are kept
  identical in order to make the comparison clearer. Furthermore, we
  applied a torque so that surface velocity at the age of the Hyades
  ($\simeq 700$ Myrs) is the same whether $E_{\mu}$ is
  zero or not.

\subsection{Meridional circulation velocity $U(r)$}

Figure~\ref{Ur} presents profiles of $U$, $\Omega$ and the 
diffusion coefficient $D_{\rm tot}$ at different times on the main
sequence, indicated in terms of central hydrogen mass fraction ${\rm X}_c$. We
superimpose the profiles obtained for $E_{\mu}$ = 0 (dashed lines) and $E_{\mu} \ne 0$ (solid lines).

When neglecting $\mu$-currents (${E}_{\mu}$ = 0), the meridional
  velocity is negative in the entire mixing
  zone. There is only one meridian loop that brings matter upwards at the equator
  and down in polar regions. This leads to the extraction of
  angular momentum in the radiative zone, a consequence of braking. The amplitude of
  meridional velocity decreases as the star evolves and braking slows down.
\begin{figure*}[t]
\centering
  \includegraphics[height=11cm,width=15cm]{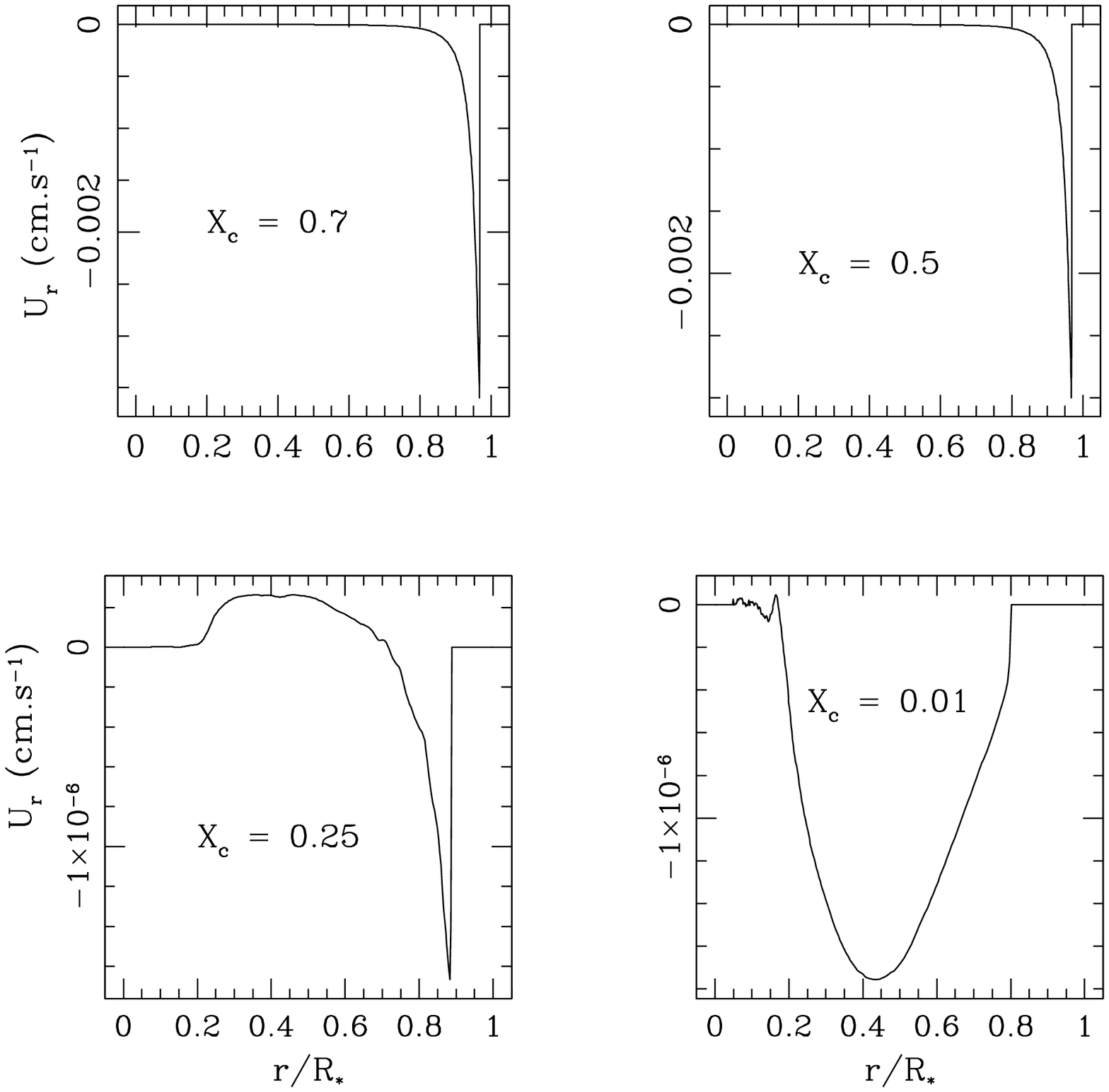}
\caption{Vertical component of the meridional circulation velocity, $U$ as a
  function of the reduced radius at the same epochs inside the same star.}
\label{EoEmu}
\end{figure*}
  
  When $E_{\mu}$ is not zero, some changes appear. 
  There are now two loops of circulation. In the outer parts of
  the radiative zone, the negative loop keeps on transporting material from
  the equator up along the pole, in response to the extraction of angular momentum. 
  In the interior however, circulation is positive, revealing an inward
  transport of angular momentum. Equations (\ref{vcirc}),~(5) and (\ref{emu})
  already suggest a strong interaction between $\Omega$- and
  $\mu$-currents. Apart from the first term in Eq.~(\ref{vcirc}), which includes
  Eddington-Sweet and Gratton-\"Opik terms, $E_{\Omega}$ and 
  $E_{\mu}$ are almost mirrored. 
  At the beginning of the evolution, $|E_{\Omega}| > |E_{\mu}|$, for
  $\mu$-currents have not settled yet. As the star evolves and $\mu$-gradients
  built-up, $\mu$-currents may grow.
  $|E_{\mu}|$ and $|E_{\Omega}|$ eventually become of the same order of magnitude,
  and $U$ disminishes.

  Contrary to what was expected by Mestel (1953) the circulation is
  not frozen. Indeed, the two terms $E_{\mu}$ and $E_{\Omega}$, while being of the same order, never
  compensate each other as can be seen in Fig.~\ref{EoEmu}, where $U$
  becomes very small, but never disappears completely. Indeed, due to the
  strength of the horizontal diffusion, the circulation must remain to
  continuously refresh the horizontal $\mu$ fluctuation. However, the
  difference between both contributions $E_{\Omega}$ and $
  E_{\mu}$  decreases significantly as the star evolves. Meridional
  circulation velocity is thus quite slow near the end of the main
  sequence, but transport of chemicals still occurs, as can be seen from
  the diffusion coefficient in Fig.~\ref{Ur}.  

\subsection{Rotation profile $\Omega$}

The inner transport of angular momentum also affects the rotation profiles
(Fig.~\ref{Ur} middle panels): when including $\mu$-currents, 
differential rotation grows larger. In particular, the core gains
momentum and gets accelerated as evolution proceeds instead of keeping a
nearly constant rotation rate. 
This can be of crucial importance when
further following the evolution of the star on the red giant branch.
This behaviour may be interpreted as follows: the rate of differential
  rotation and the relative variations of mean molecular weight are related
  to each other via the equation of state (Eq. \ref{EOS}). Observations and
  theory show us that the circulation in main sequence stars on the hot
  side of the Li dip must act to extract angular momentum. In this
  configuration, the term $E_{\Omega}$ creates an anti-clockwise circulation
  transporting matter from the pole down to the equator. $E_{\mu}$, on the
  otherside, opposing $E_{\Omega}$ (see Eq. \ref{vcirc}), generates a
  clockwise circulation. The net effect is a reduction of angular
  momentum transfer to the surface and an enhancement of differential rotation.

\begin{figure}[t]
\resizebox{\hsize}{!}{\includegraphics[angle=-90]{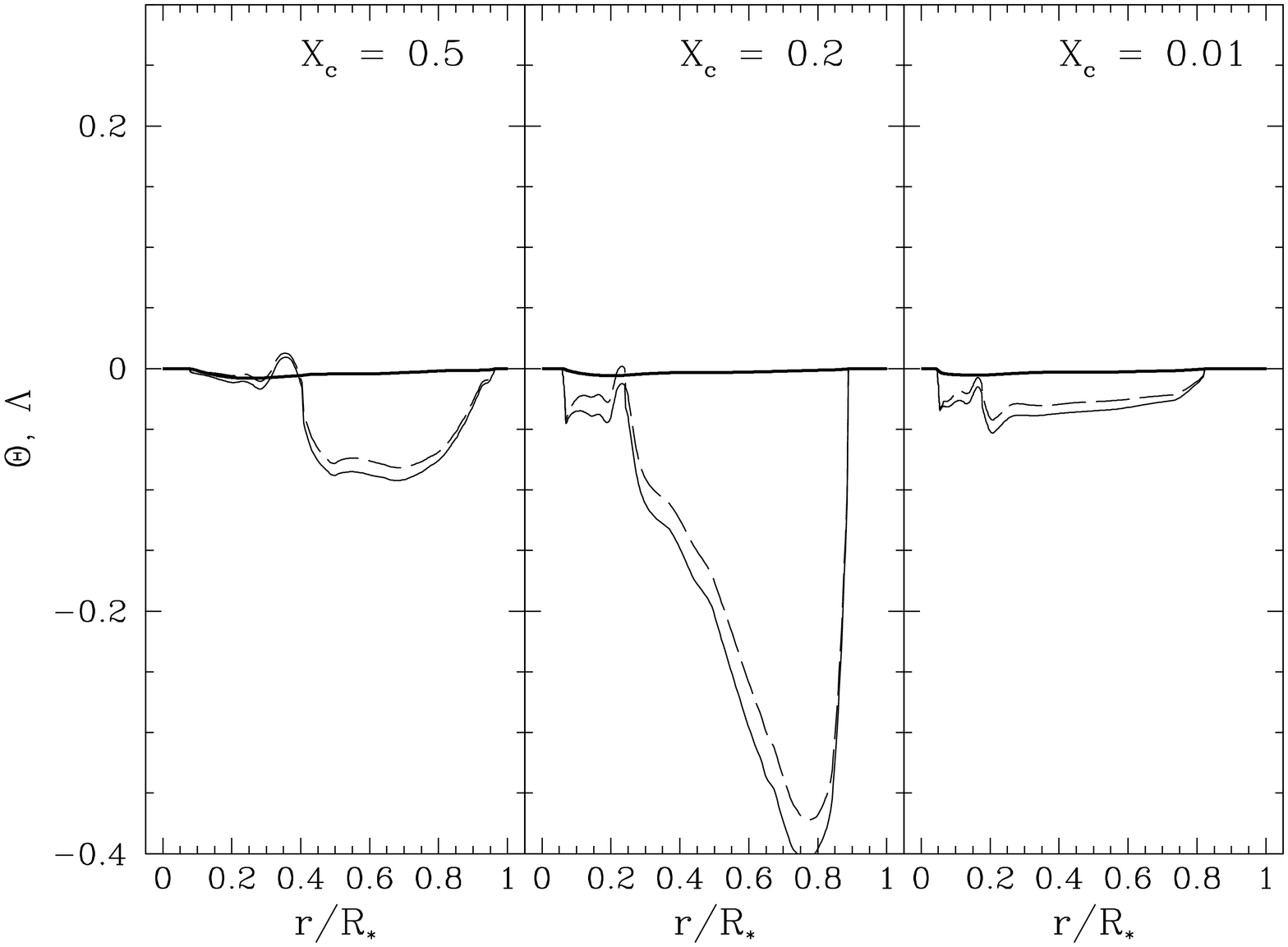}}
\caption{Profiles of $\Theta$ and $\Lambda$ inside the 1.5 ${\rm M}_{\odot}$
  star for three different values of the central hydrogen content. Solid
  lines are for $\Theta$, dashed ones for $\Lambda$. Thin lines stand for
  models where ``complete mixing'' (${\rm E}_{\mu} \neq 0$) is applied
  while the bold line gives the profile of $\Theta$ when ${\rm E}_{\mu} =
  0$.}  
\label{om}
\end{figure}

In Fig.~\ref{om} we present profiles of $\Theta$
and $\Lambda$ at given evolutionary points. The bold line represents the
$\Theta$ profile in the absence of $\mu$-currents. The difference between
both is striking. It shows, as was already pointed out by Talon et
al. (1997) in their study of a 9~${\rm M}_{\odot}$ star, that the
description of the rotation profile and of the circulation is not anymore a
matter of $\Theta$ (or $\Omega$) alone, but of $\Theta - \Lambda$. 
Indeed, $\Theta$ is then conditioned by $\Lambda$ due again to their
mirrored evolution (Eqs.~5 and \ref{emu}).

The large central differential rotation is a key ingredient in core
evolution. Indeed, $\mu$-gradients in that region inhibit the development
of a large turbulent diffusivity (see Eq.~\ref{lambda}). However, the
large shear resulting from the mirror evolution of $\Theta$ and $\Lambda$
significantly increases local core mixing\footnote{As explained in Talon et al. 1997,
in a non-evolving model, Eq. (2) has a stationnary solution in which advection
of momentum by the circulation is compensated by turbulent transport. There is
thus a $\Theta$ equilibrium profile. When adding $\mu$-currents, about the same
equilibrium exists for $\Theta - \Lambda$. This leads to $\Theta$ following the
behavior of $\Lambda$, resulting in large differential rotation
in the core.}.

Let us mention the work of Th\'eado \& Vauclair (2001), who performed
calculations including the effects of $\mu$-currents according to
Vauclair (1999) in order to reproduce the low dispersion found in the
``Spite lithium plateau''.
Assuming slow, constant, solid body
rotation (a few km.${\rm s}^{-1}$ in typical halo dwarfs),
and that the transport of chemicals is progressively blocked when $E_{\mu} = E_\Omega$
(a situation they call {\it creeping paralysis}),
they achieve a dispersion of 0.1 dex in the lithium plateau,
in agreement with observations.
We point out the fact that their scenario implies solid body rotation and
thus prevents the rotation profile to react to horizontal molecular
weight gradients as is the case in our description. Solid body rotation
would be enforced in the presence of a small magnetic field or by gravity waves
(Talon \& Charbonnel, in preparation), which might be the
case in halo stars as well as in the stars lying on the cool side of the Li
dip. However, as discussed by TC98, such an additional
process of extraction of angular momentum is not expected to be efficient in the 
more massive stars lying on the left side of the Li dip which are under the
scope of this study.

\subsection{Diffusion coefficients}

\begin{figure}[t]
\resizebox{\hsize}{!}{\includegraphics{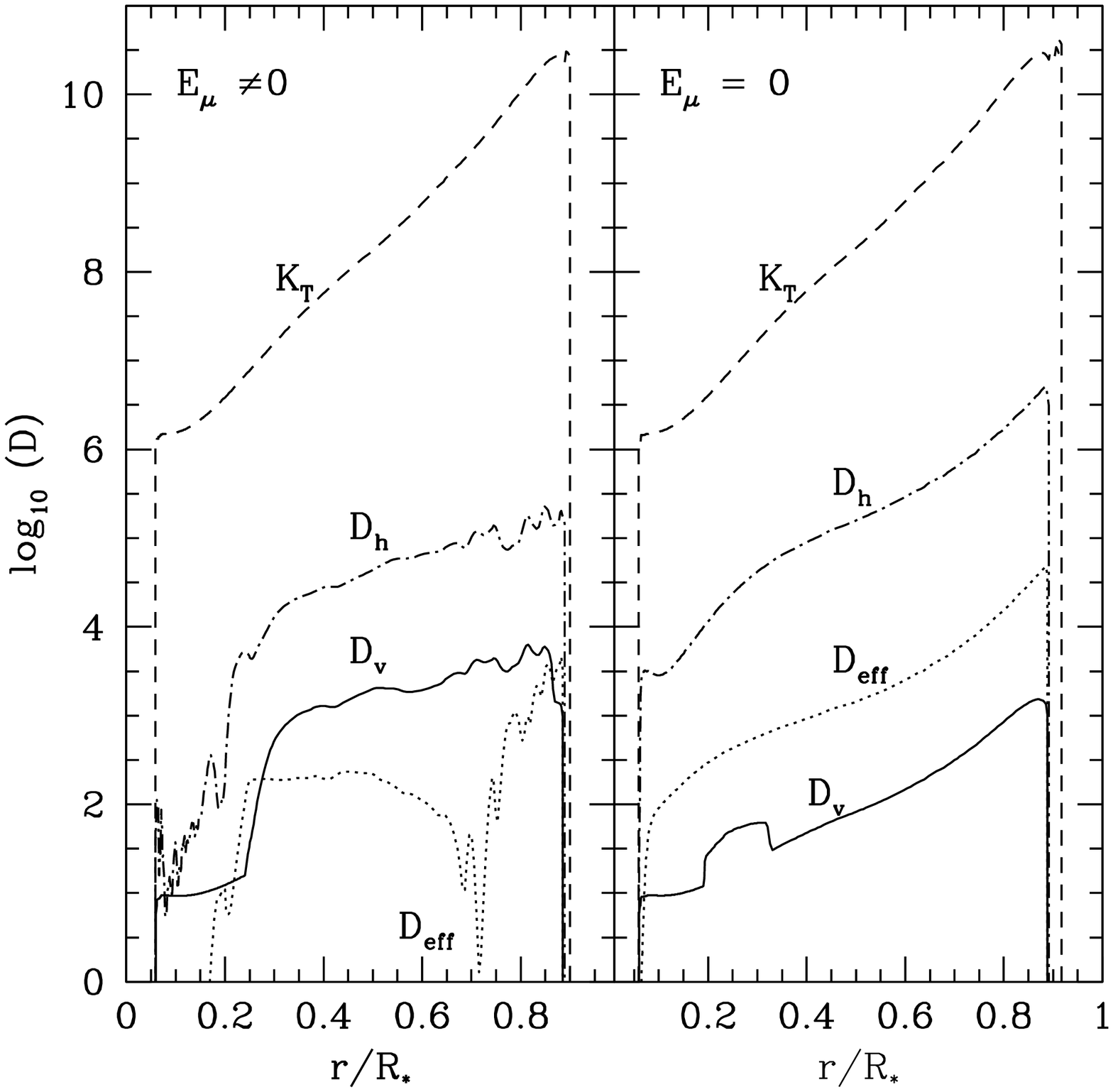}}
\caption{Profiles of the different components of the diffusion coefficient
  and of the thermal diffusivity as a function of the reduced radius when
  $E_{\mu} \ne 0$ (left) and $E_{\mu} = 0$ (right). The
  models correspond to the $1.5 {\rm M}_{\odot}$ star with ${\rm X}_c = 0.2$. }  
\label{Dprof}
\end{figure}

The examination of the total diffusion coefficients (from Fig.~\ref{Ur}), 
reveals that, except towards the centre, they appear to be very much the
same, whether $E_{\mu}$ is zero or not. The transport of chemicals, contrary to that of
angular momentum, will thus not be significantly affected during main
sequence evolution (see next section). 
However, while the total coefficients are of similar amplitude, they are not
the result of the same combination of effects. Fig.~\ref{Dprof} presents
various diffusion coefficients in both cases for a central hydrogen
content ${\rm X}_c = 0.2$. 
We can firstly emphasize the fact that the shellular
rotation is actually realistic, for
$D_h \gg D_{\rm eff} \; {\rm
  and} \; D_v$. 
Secondly, when $E_{\mu} \ne 0$, the turbulent diffusion
coefficient $D_v$ is larger than the effective coefficient $D_{\rm eff}$ almost
everywhere in the radiative zone. This is a consequence of the
  enhancement of the differential rotation rate in this case, which favours
  the growth of turbulence\footnote{This appears clearly when turning to
    Eq. (\ref{Dv}), through which $D_v \propto {(d \Omega /dr)}^2$.}. Due to the strong anisotropy of turbulent
motions, the buoyancy restoring forces in the vertical direction are
slightly reduced because of the horizontal diffusion, and vertical
turbulent mixing is allowed. The profile of the effective diffusion
coefficient is influenced by $U(r)$ (Eq.~\ref{Deff}), so that it reflects
the loops of the circulation in the presence of $\mu$-currents, each
``drop'' corresponding to an inversion of the direction of the flow. 

When neglecting variations of $\mu$, turbulence can only
marginally develop in a small portion of the radiative zone and the vertical
diffusion coefficient $D_v$ is
reduced to molecular viscosity which is always smaller than
$D_{\rm eff}$. 

In the central part, we may notice that the total diffusion coefficient is
lowered by one order of magnitude in the presence of $\mu$-currents. As
pointed out before, $\mu$-gradients in that region inhibit the development
of an efficient turbulent diffusivity. This appears clearly in
Fig.~\ref{Dprof} (left panel), where we see $D_{\rm eff}$ dropping down
below $10^1$ just above the core. $D_v$ also decreases in the same
region, where contrary to the rest of the radiative zone, it is dominated
by the molecular viscosity rather than by the turbulent viscosity (there is
no turbulence).  
\begin{figure}[t]
\resizebox{\hsize}{!}{\includegraphics{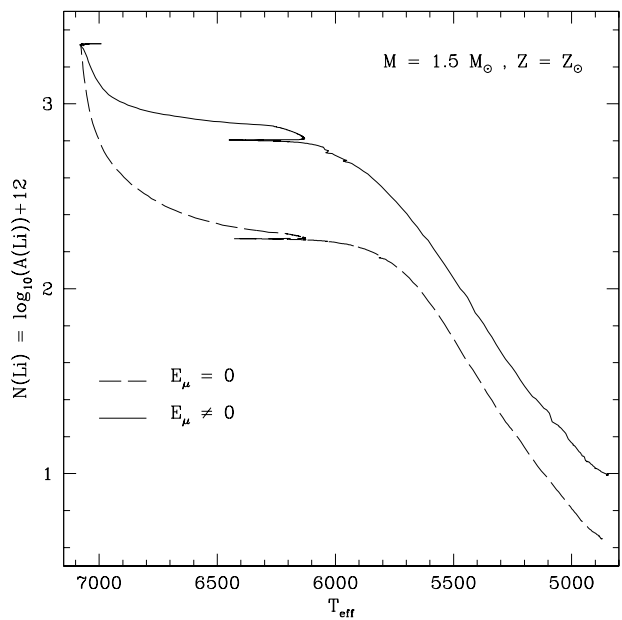}}
\caption{Evolution of the surface abundance of lithium with temperature in
  the 1.5 ${\rm M}_{\odot}$ star for the two different approaches of mixing
  as mentioned on the plot and in the text.}
\label{teta}
\end{figure}

Figure~\ref{teta} presents a comparison between the evolution of the
lithium surface abundance in the $1.5 {\rm M}_{\odot}$ rotating star when considering or not the
effects of $\mu$-currents. Both tracks are similar, but the depletion
of lithium during the main sequence is larger in case of
$E_{\mu} = 0$. The surface abundance of lithium differs by more than 0.3
dex at 700 Myrs, which is more than the observational errorbar commonly
assumed. Accounting for $\mu$ variations reduces the effect of
mixing {\em when all others assumptions remain the same}. We have seen that 
$\mu$-currents rule the evolution of the rotation profile. They lead to an
enhancement of the degree of differential rotation, which in turn allows
for the horizontal turbulence to develop more easily. This horizontal
turbulence hinders the transport towards the vertical direction. This
effect shows up through the surface abundance of lithium, that is less
depleted when $E_{\mu} \ne 0$. 
Nevertheless, this abundance feature can not be advocated as an
observational clue to the action of the $\mu$-currents. Indeed, braking
plays a major role in determining the amplitude of mixing (and thus the
amplitude of lithium depletion), and thus may rub off the effects of the
variations of $\mu$. 
The surface velocity reached for ${\rm T}_{\rm eff} \simeq 4900~{\rm K}$ is of
a few ${\rm km}.{\rm s}^{-1}$ in both cases, velocity for the model with
$E_{\mu} = 0$ being slightly larger ($9~{\rm km}.{\rm s}^{-1}$ versus
$6.5~{\rm km}.{\rm s}^{-1}$ when $E_{\mu} \ne 0$), because of the different
braking constant used in this case in order to achieve the same surface
velocity at 700 Myrs as in absence of $\mu$-currents. Both velocities
reached at the end of the computation are consistent with observations
(Gaig\'e 1993).
 \begin{figure}[t]
\resizebox{\hsize}{!}{\includegraphics{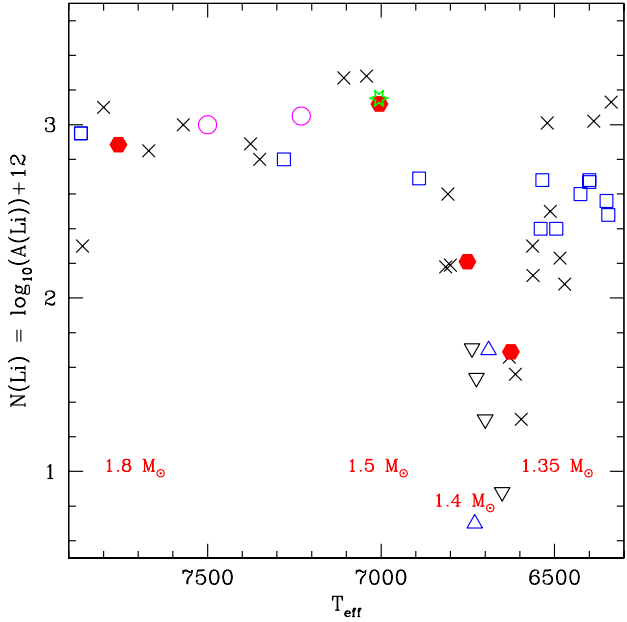}}
\caption{Comparison of the models (hexagons and open bold star) with observations in
  the Hyades (crosses), in Praesepe (squares) and in Coma Berenices
  (circles and triangles). Downward triangles are for upper limits. Observations are from Burkhart \& Coupry
  (1998,2000) and Boesgaard (1987). The corresponding masses of the models
  are as indicated on the figure. All models presented have Z = 0.02 and an initial velocity of 110 km.${\rm s}^{-1}$; the
  reader is referred to Tab.~\ref{table1} for details on braking. For
  the 1.5 M$_{\odot}$ model, open bold star is obtained
  when not including microscopic diffusion on LiBeB while for hexagons
  microscopic diffusion was included also for these light elements (see
  text for further details).}
\label{LiMS}
\end{figure}
As a result, accounting for the effects of $\mu$-currents under the
assumptions described in \S~3 does not make any significant change in the
outcoming surface abundances compared to the case without variations of
$\mu$. The observed abundances used as constraints do not allow to
distinguish between these two cases. However we pointed out differences in the 
final rotation profiles, which might change significantly rotational
transport during later evolutionary phases.

\section{Comparison with observations}
We now compare the predictions of our rotating-models with the
observations of Li, Be and CNO on the hot side of the so-called Li dip
at the age of the Hyades, and of Li in subgiants of solar metallicity.

\subsection{Li, Be and CNO on the hot side of the Li dip in the open
  clusters}

\begin{figure}[t]
\resizebox{\hsize}{!}{\includegraphics{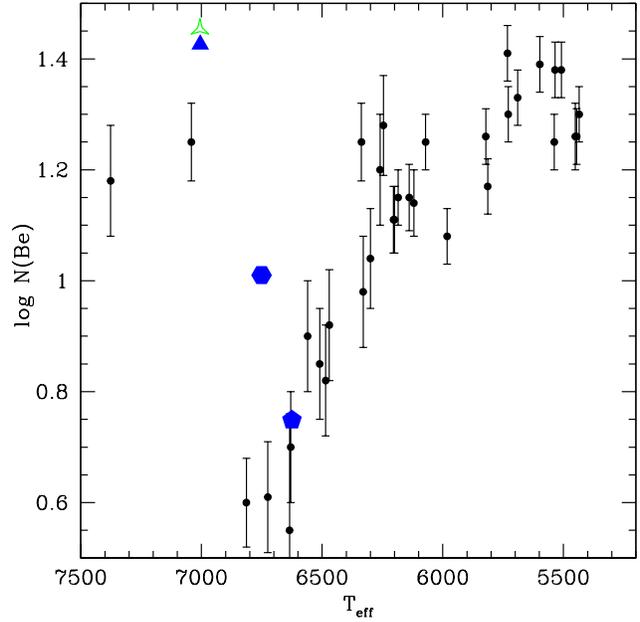}}
\caption{Beryllium abundance as a function of effective temperature
  in the Hyades (Boesgaard \& King 2002).
The pentagon, hexagon and triangle are the values obtained
for the 1.35 ${\rm M}_{\odot}$, 1.4 ${\rm M}_{\odot}$ and 1.5 ${\rm
  M}_{\odot}$ models respectively. The empty triangle is a 1.5 ${\rm
  M}_{\odot}$ model where microscopic diffusion is not applied to
LiBeB. Refer to Tab. 1 for further details on braking.}     
\label{Be}
\end{figure}

The drop-off in the Li content of main-sequence F stars in a narrow
range of effective temperature centred around 6700~K, or ``Li dip"
(Wallerstein et al. 1965, Boesgaard \& Tripicco 1986, Balachandran 1995), 
is the most striking signature of transport processes that occur 
inside these stars. 
We refer to TC98 and CT99 for a detailed discussion of the different
processes which have been proposed to account for this feature. 
These papers show how rotation induced mixing shapes in a
very natural way the hot side of the Li dip. As already mentioned, we did
not consider the effects of the $\mu$ gradients in these previous papers.
However, as discussed in \S~4.2.3, the total diffusion coefficients, and
thus the resulting transport of chemicals, is not significantly
affected by these terms during main sequence. As a consequence, 
the comparison of the predictions of Li and Be
surface abundances with observations on the hot side of the Li dip
remains highly satisfactory,  as can be seen in Figs.~\ref{LiMS} and \ref{Be}.   

In CT99, microscopic diffusion on LiBeB had been turned off (while
it was applied to all other chemicals) for stars with $T_{\rm eff}$
higher  than 6900~K, because of the competition between radiative acceleration 
and gravity (Richer \& Michaud 1993, Richer et al. 2000, see discussion 
\S~\ref{diff_micro}). 
In Figs.~\ref{LiMS} and \ref{Be}, we present the
results of computations with and without gravitational settling of
LiBeB at the effective temperature where radiative forces can
over-compensate gravitation. As can be seen, for the rotationnal velocities considered here,
the differences are minute. In addition,
the effect on stellar structure itself is null, since LiBeB are
trace elements. 

\begin{figure}[t]
\resizebox{\hsize}{!}{\includegraphics{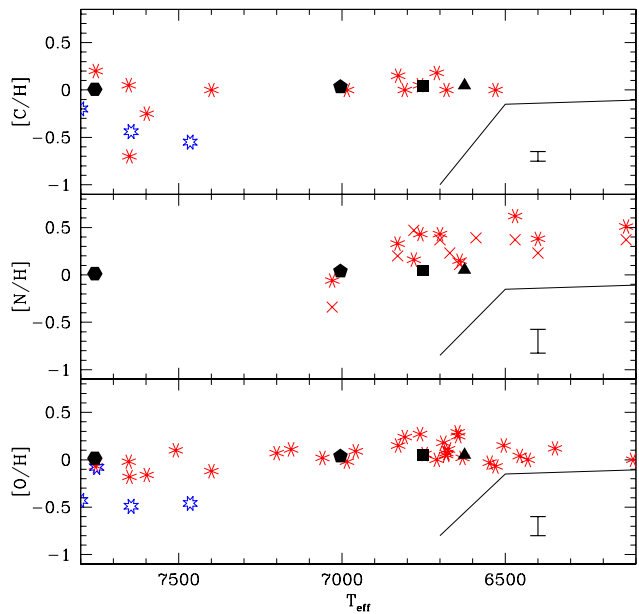}}
\caption{Abundances of carbon, nitrogen and oxygen vs effective temperature
  in the Hyades, presented as asterisks and open stars for A and Am stars
  respectively (the latter ones are not meant to be reproduced here; see
  text for details). Observations of C and O are from Varenne \& Monier
  (1998). Observations of N are from Takeda et al. (1998), crosses are
  derived from the 8629 ${\rm \AA}$ line and stars are from 8683 ${\rm
  \AA}$ line. Abundances are given relative to solar values from Grevesse
  \& Noels (1993). The typical errorbar for observational values is given
  for each panel. Models with complete mixing, a turbulence
  parameter ${\rm C}_h = 1$ and braking as reported in Tab.~\ref{table1} are
  represented by the filled points with hexagon, pentagon, square and triangle
  standing for 1.8, 1.5, 1.4 and 1.35 ${\rm M}_{\odot}$ respectively, with
  ${\rm Z}_{\star} = 0.02$. Lines are from calculations including
  gravitational settling and radiative forces by Turcotte et al. (1998)
  (see Varenne \& Monier 1999 for further details).} 
\label{CNOHyades}
\end{figure}

Observations of C, N and O are precious clues on
transport mechanisms inside main-sequence F-type stars, and on the
competition between atomic diffusion and macroscopic processes. Indeed, in
the case of pure atomic diffusion, these elements are expected to be
underabundant in the Li dip, as shown in Fig.~\ref{CNOHyades}. On
the other hand, in our models rotation induced mixing strongly inhibits
the settling effects.  As a result, the surface abundances of CNO do not
vary in the $T_{\rm eff}$ range  considered. This is in perfect agreement
with observations in the Hyades  (Varenne \& Monier 1998, Takeda 1998;
see also Boesgaard 1989, Friel \& Boesgaard 1990, Garcia Lopez et
al. 1993), as can be seen in Fig.~\ref{CNOHyades}. 
Let us stress that in the temperature domain of A-type stars, most of the
observations report abundances for chemically peculiar objects, namely Am
stars, that are slow rotators in which the action of gravitational settling
is important on CNO elements. These peculiar patterns might be achieved by
considering the combined effects of large scale circulation, turbulence,
gravitational settling and radiative forces (the effects of the latter not
being included in our evolutionary code) in models with a slow initial rotation.
Such computations are not presented here for this is out of the scope of
the present study to focus on such a particular phenomenon.

\subsection{Li in subgiants}

\begin{figure}[t]
\resizebox{\hsize}{!}{\includegraphics{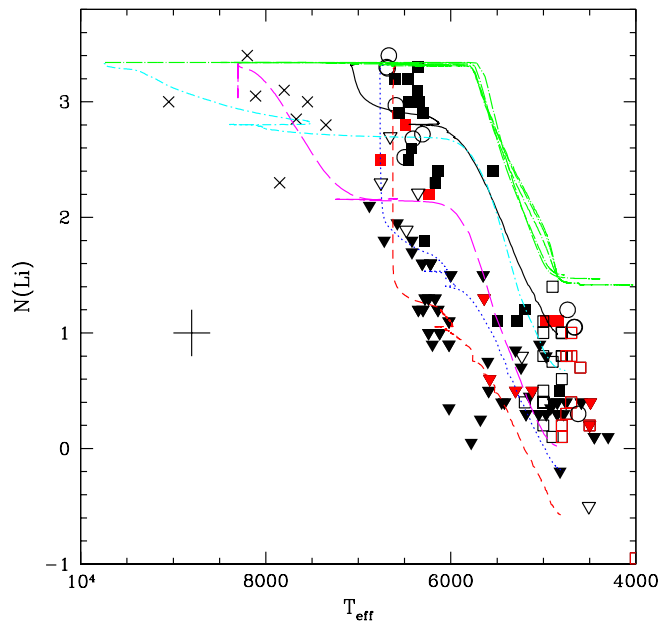}}
\caption{Lithium abundance versus effective temperature in open clusters
  (open symbols) and field (filled symbols) subgiants compared with
  our rotating models. Triangles indicate upper limits. Observations
  are from L\`ebre et al.(1999) (field), Wallerstein
  et al. (1994) (field), Pasquini et al. (2001) (NGC 3680), Gilroy
  (1989) (open clusters) and Burkhart \& Coupry
  (1989,2000) (Hyades among other open clusters). Dot-dashed lines
  represent results for standard models of $1.35 {\rm M}_{\odot}$, $1.4
  {\rm M}_{\odot}, 1.5 {\rm M}_{\odot}$, $1.8 {\rm M}_{\odot}$ and $2.2
  {\rm M}_{\odot}$. They all coalesce in a single line, standard dilution
  starting around 5750~K. Other lines are for models with
  rotation induced mixing and magnetic braking. Solid line, dotted line,
  long dashed line, dashed line and dot-dashed line are for $1.5 {\rm
  M}_{\odot}$, $1.4 {\rm M}_{\odot}$, $1.8 {\rm M}_{\odot}$, $1.35 {\rm
  M}_{\odot}$, $2.2 {\rm M}_{\odot}$ respectively. Refer to Tab.~1 for
  details concerning rotational velocities.}  
\label{SGB}
\end{figure}

Early observations by Alschuler (1975) of a few field stars crossing for
the first time the Hertzsprung gap indicated that lithium depletion starts 
earlier than predicted by standard dilution for stars more massive than
2~M$_{\odot}$.
Later on, significant lithium depletion was observed in a non negligible
number of slightly evolved field stars (Brown et al. 1989, Balachandran
1990).
do Nascimento et al. (2000) studied the behaviour of lithium in a 
sample of field subgiants observed by L\`ebre et al. (1999) for which
Hipparcos data allowed the precise determination of both evolutionary 
status and mass. They confirmed that stars originating from the hot side 
of the dip present a large range of lithium abundances which can not be 
explained by standard dilution alone and which reflect different degrees of 
depletion of this element while on the main sequence, even if its signature 
does not appear at the stellar surface at the age of the Hyades (see Vauclair
1991, Charbonnel \& Vauclair 1992).
In this cluster, dilution is not sufficient to explain lithium values in
evolved stars (which have masses of the order of 2.2~M$_{\odot}$), while
main sequence stars present galactic abundance (Boesgaard et al. 1977,
Duncan et al. 1998). This has been observed also in open clusters with 
turnoff masses higher than $\sim$ 1.6~M$_{\odot}$ (Gilroy 1989, Pasquini
et al. 2001).

In Fig.~\ref{SGB} we compare our predictions for the surface Li abundance of various 
models with the observations in both field and open cluster evolved stars. 
First note that in classical models without rotation the variation 
in surface Li abundance is very similar for stars of different
masses because it is only due to the first dredge-up dilution. More
importantly, classical Li depletion starts at a too low 
$T_{\rm eff}$ (around $\sim 5850$~K) and 
leads to final Li abundances too high when compared to observational data.
On the other hand, and even if no important lithium depletion is
expected at the surface of the main sequence stars on the hot side of the 
dip at the age of the Hyades, more lithium destruction occurs inside 
rotating models compared to classical ones. 
Indeed, due to the enlargement of the Li free radiative regions, important surface Li
depletion occurs at much higher $T_{\rm eff}$ than in classical models. Also,
total Li depletion after dredge-up is more important in 
rotating models than in models without transport processes. 
The predicted evolution of the surface lithium abundance in our rotating
models explains the behaviour and the dispersion in field and open cluster 
evolved stars more massive than $\sim$~1.4~M$_{\odot}$. 

\section{Conclusions}

To study the evolution of low mass rotating stars and the
effects of rotation induced mixing on surface abundance patterns, we have
presented here a first series of results concerning main sequence and
subgiant Pop I stars located on the hot side of the lithium dip.
We performed calculations including the treatment of transport of angular
momentum and chemical species due to meridional circulation, shear
turbulence and gravitational settling according to Maeder \& Zahn (1998)
formalism. In particular, we take into account horizontal variations of mean
molecular weight, and their action on the transport itself.
These terms prove to be very important in defining the shape of the
internal rotation profile. Indeed the horizontal variations of $\mu$ and
the derivative of $\Omega$, namely, $\Lambda$ and $\Theta$ respectively,
have a mirrored behaviour, $\Theta$ following $\Lambda$ when it is not
null. As soon as the circulation is settled, the $\mu$-currents grow larger
and induce a stronger differential rotation.

On the other hand, the effects of these terms on the surface abundances of
light elements appear not to be constraining compared with braking for instance. The major change introduced in the
transport of chemicals when taking $\mu$-currents into account concerns the
dominant contributor to the diffusion coefficient. In an ``homogeneous'' star
(where there are no horizontal variations of $\mu$), the effective
diffusion coefficient $D_{\rm eff}$ is the major contributor, while in the
inhomogeneous case, turbulent diffusion coefficient dominates the transport
of chemicals.

Finally, contrary to what has been suggested in previous work concerning
rotation induced mixing in solid body rotating low mass stars (typically
halo main sequence stars with ${\rm M}_{\star} < 0.9 {\rm M}_{\odot}$), when
considering extraction of angular momentum through magnetic braking, the
situation of ``{\it creeping paralysis}'' of the circulation is not reached
in the models presented here. It shall actually never be the case
as long as differential rotation is allowed to exist
for the mirrored terms never compensate each other.  

The point to be emphasized is the fact that these variations of mean
molecular weight are not to be neglected when considering rotating
objects. They modify the action of mixing and can be of major importance
after main sequence evolution, for they constrain the internal rotation profile.

\begin{acknowledgements}
We are grateful to the referee Pr. Andr\'e Maeder and Pr. Sylvie
  Vauclair for their usefull comments, that helped to improve this paper.
We wish to thank the french ``Programme National de Physique Stellaire''
for financial support. S.T. was supported by NSERC of Canada and by the Canada Research
Chair in Stellar Astrophysics awarded to G.~Fontaine. A.P. would like to
  thank the Centre Informatique National de l'Enseignement Sup\'erieur
  where calculations were performed. 
\end{acknowledgements}

\appendix

\end{document}